\documentclass[10pt]{iopart}

\usepackage[T1]{fontenc}
\usepackage[utf8]{inputenc}
\usepackage{graphicx}
\usepackage{tabularx}
\usepackage{color, soul}
\usepackage{textcomp}
\usepackage{enumerate}
\usepackage{todonotes}
\usepackage[draft]{hyperref}
\usepackage{comment}
\usepackage{subcaption}
\usepackage{url}
\usepackage[normalem]{ulem}
\usepackage{makecell}
\usepackage{graphbox}
\usepackage{stfloats}
\usepackage{booktabs} 
\usepackage{multirow}
\usepackage{cite}
\usepackage{amssymb}

\begin{document}

\title[Investigation of Inverse Bremsstrahlung Heating Driven by Broadband Lasers]{Investigation of Inverse Bremsstrahlung Heating Driven by Broadband Lasers} 

\author{Xiaoran Li$^{1}$, Jie Qiu$^{1}$, Liang Hao$^{1}$\footnote{Corresponding author: hao\_liang@iapcm.ac.cn}, Chen Wang$^{2}$\footnote{Corresponding author: wch11@163.com}, Lifeng Wang$^{1}$, Shiyang Zou$^{1}$}

\address{
	$^1$ Institute of Applied Physics and Computational Mathematics, Beijing 100094, People’s Republic of China}
\address{$^2$ Shanghai Institute of Laser Plasma, CAEP, Shanghai 201899, People’s Republic of China}


\vspace{10pt}
\begin{indented}
        \item[]\today
\end{indented}

\begin{abstract}
	Broadband lasers have become a key strategy for mitigating laser plasma instabilities in inertial confinement fusion, yet their impact on collisional inverse bremsstrahlung (IB) heating remains unclear.
	Using one dimensional collisional particle-in-cell simulations, we systematically examine the effect of bandwidth induced temporal incoherence on IB absorption in Au plasmas.
	The simulations are first benchmarked against classical absorption theory, verifying that the implemented Coulomb collision model accurately reproduces the theoretical IB heating rate.
	{A direct comparison of the electron temperature evolution in the broadband and monochromatic cases shows that, although spectral broadening introduces transient picosecond scale oscillations in the heating rate driven by stochastic intensity fluctuations, the long term averaged heating and net IB absorption remain essentially unchanged.
	}
	
	
\end{abstract}

\begin{indented}
\item[] \textbf{Keywords}: laser driven inertial confinement fusion, broadband lasers, inverse bremsstrahlung heating
\end{indented}

\ioptwocol

\section{Introduction}

Broadband lasers have recently attracted increasing attention in inertial confinement fusion (ICF) as a promising approach to mitigate laser plasma instabilities (LPI).
By introducing frequency spread and temporal incoherence, broadband laser disrupts the phase matching among resonant waves and effectively suppresses the coherent amplification of parametric instabilities.
Recent experiments on both the Kunwu and PHELIX laser facilities have demonstrated effective mitigation of {SBS in CH targets} with a modest bandwidth of 0.5\%--0.6\%~\cite{kang_effects_2025, lei_reduction_2024, wang_backward_2024, kanstein_2025}, suggesting that spectral broadening can play a crucial role in achieving LPI-free ignition schemes.
{Meanwhile, extensive computational studies have further examined how processes such as stimulated Raman scattering (SRS), two plasmon decay (TPD), stimulated Brillouin scattering (SBS), and cross beam energy transfer (CBET) depend on the bandwidth and intensity of broadband lasers, thereby extending our understanding of broadband driven laser plasma interactions~\cite{zhao_control_2024, liu_parametric_2023, bates_suppressing_2023, yin_mitigation_2025, seaton_cross-beam_2022, li_2025_sbs}.}

{In contrast, a closely related question has received far less attention: whether introducing laser bandwidth also alters the collisional inverse bremsstrahlung (IB) absorption that dominates energy deposition in the subcritical plasmas relevant to ICF.}  
Since IB heating depends on both the instantaneous laser intensity and the evolving plasma state~\cite{michel2023}, the stochastic sub-picosecond fluctuations inherent to broadband light could, in principle, could modify the temporal structure of collisional energy deposition even when the long time mean intensity is fixed.  
This motivates a central question: does spectral broadening intrinsically change the physics of IB heating, or does it simply modulate the instantaneous intensity that drives an otherwise identical absorption process?  
Clarifying this issue is essential for evaluating the extent to which broadband driven plasmas differ from monochromatic ones in both kinetic and fluid descriptions.

In this work, we use collisional particle-in-cell (PIC) simulations to isolate the role of bandwidth induced temporal incoherence on IB heating, keeping the mean intensity, central frequency, and plasma parameters fixed.
We first verify that the collisional absorption computed by the PIC model is consistent with classical theory, confirming the correct implementation of the Coulomb collision operator.
We then compare the temporal evolution of the electron temperature $T_e$ and the heating rate $dT_e/dt$ under monochromatic and broadband illumination.
The results show that while the broadband laser produces short time oscillations in the heating rate due to its stochastic intensity modulation, the long term evolution and overall absorption efficiency remain nearly identical to those of the monochromatic case.

\section{Simulation Model and Parameters}

\subsection{Simulation setup}
\label{sec:domain}

The influence of laser bandwidth on the energy absorption and electron heating of high-Z plasmas
was investigated using the one-dimensional, three-velocity (1D3V) particle-in-cell (PIC) code \textsc{EPOCH}~\cite{arber2015}.
Binary Coulomb collisions among all species were included through the Nanbu Monte Carlo algorithm~\cite{nanbu1998,sentoku2008},
allowing for self consistent modeling of collisional absorption.

The simulated target was a spatially uniform Au plasma with an electron density of $n_e = 0.1n_c$,
where $n_c$ is the critical density corresponding to a laser wavelength of 530~nm.
The initial electron and ion temperatures were $T_e = 200~\mathrm{eV}$ and $T_i = 10~\mathrm{eV}$, respectively.
The Au charge state was fixed at $Z = 50$, and the ion density was set to $n_{\mathrm{Au}} = n_e / 50$ to ensure quasineutrality.
Each cell contained 10,000 electron and 200 ion macroparticles with equal particle weights, which was found to improve the accuracy of electron--ion collisional energy exchange.
The total simulation domain was $L = 300\,c/\omega_0$, which was discretized into 4500 spatial cells, corresponding to a resolution of $\Delta x \approx 0.067\,c/\omega_0$, where $\omega_0$ is the laser frequency with a wavelength of 530~nm.
Two convolutional perfectly matched layers~\cite{roden2000}, each consisting of 80 grids, are applied at both ends of the domain to absorb outgoing electromagnetic waves.  
The laser, with an intensity of $I_0 = 5\times10^{14}\,\mathrm{W/cm^2}$,
was injected from the left boundary and propagated into the plasma with open field and particle boundaries at both ends.
The total simulation time was 35~ps, sufficient to capture both the transient and steady state heating phases.

Two types of laser fields were considered:
(1)~a monochromatic laser with a wavelength of 530~nm, and
(2)~a broadband laser centered at 530~nm with a relative bandwidth of 0.6\%.
The broadband field was modeled using the multi-frequency beamlet approach,
where $N = 100$ discrete frequency components with equal amplitudes and random initial phases
were superposed to reproduce a flat top spectrum.
This approach has been widely adopted in broadband laser–plasma interaction studies
and has been shown to accurately reproduce temporal incoherence while remaining computationally efficient~\cite{liu_non-linear_2022, yao_anomalous_2024, zhao_mitigation_2024}.

The main diagnostic of interest is the spatially averaged electron temperature $T_e(t)$,
which characterizes the cumulative effect of laser energy absorption. 
The diagnostic region is chosen as the central portion of the plasma ($100$--$200\,c/\omega_0$),
where boundary effects and field reflections are negligible.
Since the IB heating rate in the present regime 
exceeds the rate of collisional thermal relaxation, 
the velocity distribution naturally deviates from a Maxwellian 
and exhibits a super-Gaussian form—an expression of the Langdon effect~\cite{langdon1980,qiu2021b}.
For each recorded snapshot, the electron velocity distribution function is extracted 
and fitted with a super-Gaussian profile
\[
f(v) \propto \exp\!\left[-(v/v_{\mathrm{th}})^{m}\right],
\]
from which the effective temperature $T_e$ is derived. 
Because the primary focus of this study is the temporal evolution of $T_e$ rather than 
the variation of the exponent $m$, the fitting parameter is fixed at $m=2.6$, 
which corresponds to the value obtained from fitting the electron velocity distribution 
at the final stage of the simulation.
Within the total simulation duration of $35\,\mathrm{ps}$, 
the electron temperature is evaluated every $\Delta t = 10^4/\omega_0 \approx 2.8\,\mathrm{ps}$ 
from successive distribution snapshots, 
allowing a detailed comparison between broadband and monochromatic laser heating dynamics.

\subsection{Verification of collisional inverse–bremsstrahlung (IB) absorption}

To validate the reliability of the collisional heating model, we benchmarked the
inverse bremsstrahlung (IB) absorption obtained from our \textsc{EPOCH} runs against
the classical collisional absorption theory.

For a homogeneous plasma with electron--ion collision frequency \(\nu_{ei}\),
the group velocity of the pump is \(v_g=c\sqrt{1-n_e/n_c}\), and the
collisional attenuation coefficient reads~\cite{michel2023}
\begin{equation}
\label{eq:absorption}
	\mu_c \;=\; \nu_{ei}\,\frac{n_e/n_c}{v_g}.
\end{equation}
Assuming Beer--Lambert decay of the pump envelope, the pump intensity varies as
\(I(x)=I_0\exp(-\mu_c x)\). The transmitted intensity after a propagation distance \(L\)
is \(T=\exp(-\mu_c L)\), giving the theoretical absorption
\begin{equation}
	A_{\mathrm{th}} \;=\; 1 - \exp\!\left(-\mu_c L\right).
\end{equation}
The collision frequency \(\nu_{ei}\) used to generate the theory curve is computed
as~\cite{nrl_formulary},
\begin{equation}
\label{eq:ei_collision}
	\nu_{ei} \;=\;
	\frac{4\sqrt{2\pi}\,n_i Z^2 e^4 \ln\Lambda}
	{(4\pi\varepsilon_0)^2\,3\,m_e^{1/2}\, T_e^{3/2}},
\end{equation}
where \(n_i\) and \(Z\) are the ion density and charge state, \(T_e\) is the electron temperature,
and \(\ln\Lambda\) is the Coulomb logarithm.

In simulations, the IB absorption \(A_{\mathrm{PIC}}\) is obtained from the spatial
evolution of the pump envelope. Let \(E_y(x,t)\) be the transverse field; its
cycle averaged envelope \(E_{\mathrm{env}}(x)\) is extracted via
Hilbert transform. To avoid boundary effects, the upstream and downstream
diagnostic windows are placed well inside the plasma, covering
\(x/L\in[0.05,\,0.15]\) and \(x/L\in[0.85,\,0.95]\).
The absorption is computed directly from the averaged field amplitudes in the
upstream and downstream regions as
\begin{equation}
	A_{\mathrm{PIC}} \;=\;
	1 - 
	\frac{\big\langle E_{\mathrm{env}}^2\big\rangle_{x\in[0.85L,\,0.95L]}}
	{\big\langle E_{\mathrm{env}}^2\big\rangle_{x\in[0.05L,\,0.15L]}}.
\end{equation}
Time averaging over many optical cycles is used to reduce statistical noise.
Note that a constant incident laser intensity is used in this test.

For a consistent comparison, the theoretical absorption is evaluated
over the same effective propagation distance as used in the PIC diagnostic.
Since the mean positions of the two windows correspond approximately to
\(x_{\mathrm{in}}\simeq0.1L\) and \(x_{\mathrm{out}}\simeq0.9L\),
the theoretical absorption is therefore given by
\begin{equation}
	A_{\mathrm{th}}^{(\mathrm{win})}
	= 1 - \exp\!\left[-\mu_c\,(x_{\mathrm{out}}-x_{\mathrm{in}})\right]
	= 1 - \exp\!\left(-\mu_c\,0.8L\right),
\end{equation}
which represents the expected attenuation between the same effective regions
used in the PIC diagnostic.

\begin{figure}[htb]
	\centering
	\includegraphics[width=0.48\textwidth]{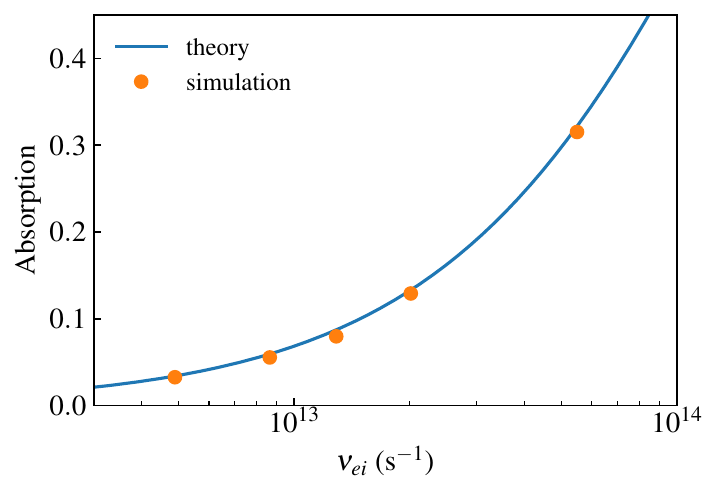}
    \caption{
	Comparison between the simulated and theoretical inverse bremsstrahlung (IB) absorption.
	The simulated absorption \(A_{\mathrm{PIC}}\) obtained from \textsc{EPOCH} is plotted
	against the theoretical prediction \(A_{\mathrm{th}}\) as a function of the electron--ion
	collision frequency \(\nu_{ei}\).
}
	\label{fig:IB_verification}
\end{figure}

Figure~\ref{fig:IB_verification} compares the simulated absorption \(A_{\mathrm{PIC}}\) with the
theoretical prediction \(A_{\mathrm{th}}\) over a wide range of electron--ion collision frequencies
\(\nu_{ei}\).
In the simulations, \(\nu_{ei}\) is varied by adjusting the electron temperature \(T_e\),
using the same plasma configuration and domain setup described in
section~\ref{sec:domain}.
The corresponding theoretical values \(A_{\mathrm{th}}(\nu_{ei})\) are evaluated using identical
parameters for direct comparison.
The test cases cover an electron temperature range of \(T_e = 200-1500~\mathrm{eV}\),
which fully encompasses the heating conditions investigated in this work.
The \textsc{EPOCH} results exhibit excellent agreement with the theoretical curve,
demonstrating that collisional energy absorption is accurately captured in the
simulations.

\section{Simulation results and discussion}

\subsection{Temporal intensity modulation of broadband lasers}

\begin{figure}[htb]
	\centering
	\includegraphics[width=0.48\textwidth]{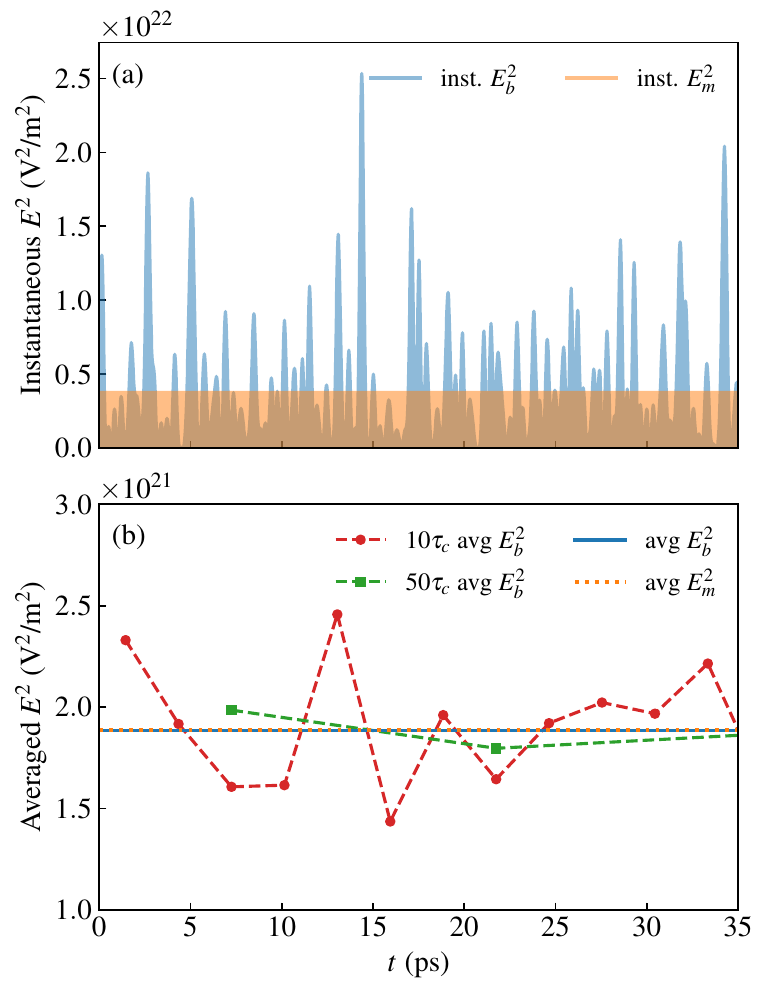}
	\caption{
		Temporal evolution of the instantaneous and averaged $E^2$ for broadband ($E_b$) and monochromatic ($E_m$) lasers at equal mean intensity, illustrating how different averaging windows ($10\tau_c$ and $50\tau_c$) smooth the broadband intensity fluctuations.
	}
	\label{fig:E2_waveform}
\end{figure}

Since inverse bremsstrahlung depends on the instantaneous laser intensity,
understanding how the broadband field fluctuates in time provides the basis for interpreting its heating behavior.
Figure~\ref{fig:E2_waveform} illustrates the temporal evolution of the instantaneous and time averaged
electric field intensity $E^2$ for the monochromatic ($E_m$) and broadband ($E_b$) lasers at the same mean intensity of
$I_0 = 5\times10^{14}~\mathrm{W/cm^2}$.
As shown in figure~\ref{fig:E2_waveform}(a), the monochromatic laser exhibits a nearly constant envelope,
whereas the broadband laser consists of a sequence of short, randomly varying wave packets,
whose instantaneous intensity fluctuates strongly in time.
The duration of these wave packets corresponds approximately to the laser coherence time,
which characterizes the timescale over which the field remains phase correlated.
{The coherence time of a broadband laser with a fractional bandwidth of
	$\Delta\omega/\omega_0 = 0.6\%$ can be estimated as~\cite{mandel1962, liu_non-linear_2022}
	\[
	\tau_c \simeq \frac{2\pi}{\Delta\omega}
	= \frac{2\pi}{0.006\,\omega_0}
	\approx 0.29~\mathrm{ps},
	\]
where $\omega_0$ is the central frequency corresponding to $\lambda_0 = 0.53~\mathrm{\mu m}$.
Each intensity spike in figure~\ref{fig:E2_waveform}(a) therefore represents a coherence burst lasting	on the order of $\tau_c \!\approx\! 0.29$~ps.}

{
	Although the instantaneous broadband intensity fluctuates strongly on the
	sub-picosecond scale, its time averaged value converges to that of the
	monochromatic laser once the averaging window significantly exceeds $\tau_c$, as
	shown in figure~\ref{fig:E2_waveform}(b).
	Averaging over a relatively short window of
	$\Delta t = 10\tau_c \approx 2.9$\,ps still leaves noticeable residual
	oscillations, reflecting partial temporal correlation across several coherence
	bursts.  
	When the averaging window is extended to
	$\Delta t = 50\tau_c \approx 14.5$\,ps, these residual modulations are strongly
	suppressed and the averaged intensity becomes nearly indistinguishable from the
	monochromatic level.
	Consistently, the broadband $E^2$ averaged over the full 35\,ps simulation is
	almost identical to that of the monochromatic case, confirming that the mean
	intensity is preserved despite rapid sub-picosecond fluctuations.  
	For a fixed averaging window, increasing the bandwidth further shortens
	$\tau_c$, thereby increasing the number of uncorrelated wave packets sampled
	within $\Delta t$ and producing an even smoother and more stable averaged
	intensity profile.
}


\subsection{IB heating dynamics with monochromatic and broadband lasers}
\label{sec:Te_evolution}

\begin{figure}[htb]
	\centering
	\includegraphics[width=0.5\textwidth]{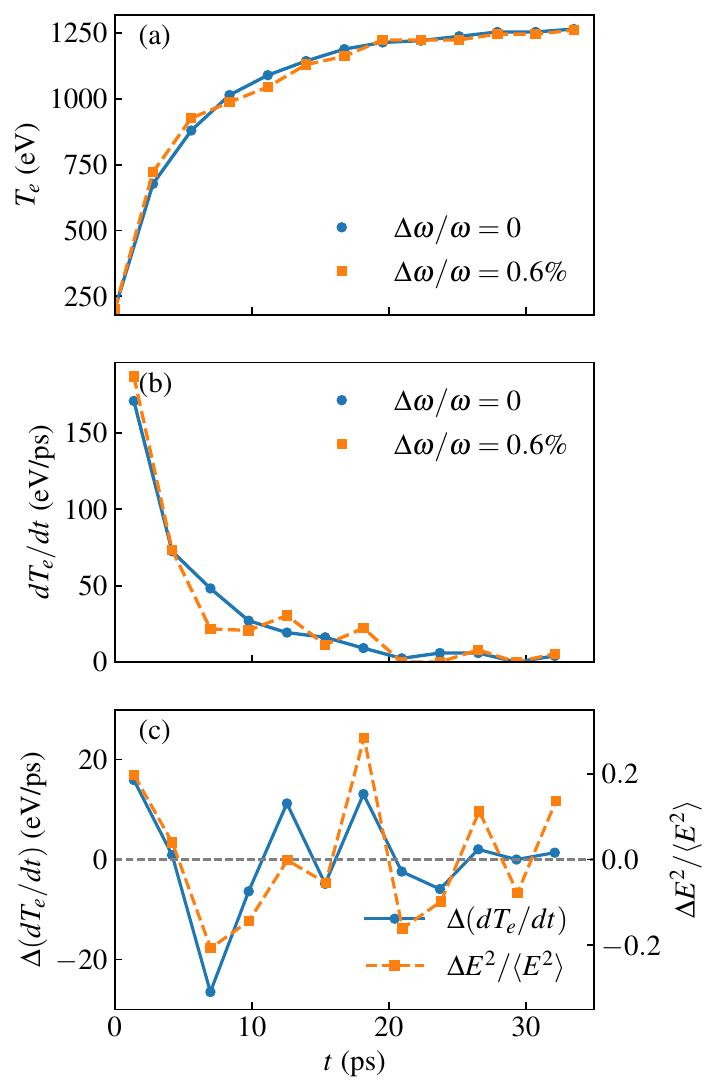}
	\caption{
		Temporal evolution of (a) the electron temperature $T_e$, 
		(b) its temporal growth rate $dT_e/dt$, 
		and (c) the difference in heating rate between broadband and monochromatic cases 
		$\Delta(dT_e/dt)$ compared with the normalized laser intensity variation 
		$(E^2_{\mathrm{b}}-\langle E_\mathrm{m}^2\rangle)/\langle E_\mathrm{m}^2\rangle$.
		Here $T_e$ is evaluated every $\Delta t = 10^4/\omega_0 \approx 2.8\,\mathrm{ps}$, $(E^2_{\mathrm{b}}-\langle E_\mathrm{m}^2\rangle)/\langle E_\mathrm{m}^2\rangle$ is estimated with the same time window.
	}
	
	\label{fig:Te_vs_t}
\end{figure}

Figure~\ref{fig:Te_vs_t} shows the temporal evolution of the electron temperature $T_e$ and its growth rate $dT_e/dt$ in a homogeneous Au plasma irradiated by monochromatic and broadband lasers.
Throughout the simulations, no significant laser–plasma instabilities such as SRS, TPD, or SBS are observed, implying that electron heating is governed almost entirely by IB absorption.

For both the monochromatic and broadband cases, the electrons absorb laser energy through IB, causing the temperature $T_e$ to rise rapidly at early times before gradually approaching a quasi saturated value of approximately $1.25~\mathrm{keV}$, as shown in figure~\ref{fig:Te_vs_t}(a). 
As $T_e$ increases, the IB absorption coefficient decreases due to the reduction of the electron--ion collision frequency, which leads to a continuous decline in the heating rate $dT_e/dt$, illustrated in figure~\ref{fig:Te_vs_t}(b). 
At sufficiently high temperatures, the effective absorption rate becomes negligible, and the plasma approaches local thermal equilibrium.

Although the overall trends of $T_e(t)$ are similar for the broadband and monochromatic cases, noticeable differences arise on picosecond time scales.
During the first $\sim6$~ps, the broadband laser produces a slightly higher electron temperature than the monochromatic one.
Afterward, the broadband $T_e$ temporarily falls below the monochromatic curve before both gradually converge at later times.
The monochromatic heating rate $dT_e/dt$ decreases almost monotonically throughout the simulation, whereas the broadband case exhibits distinct oscillations in $dT_e/dt$ despite following the same overall downward trend.
To clarify the relation between the $dT_e/dt$ fluctuations and the temporal intensity modulation of the broadband field, we compare the difference in heating rate between the two cases, $\Delta(dT_e/dt)$, with the normalized laser intensity variation 
$(E^2_{\mathrm{b}}-\langle E_\mathrm{m}^2\rangle)/\langle E_\mathrm{m}^2\rangle$, as shown in figure~\ref{fig:Te_vs_t}(c).
Here, $T_e(t)$ is sampled every $10^4/\omega_0 \approx 2.8$~ps, and the intensity variation is evaluated on the same 2.8~ps window for direct comparison.
The two curves exhibit nearly synchronous oscillations, demonstrating that the transient modulation of $dT_e/dt$ directly follows the time averaged fluctuations of the laser intensity.
	
To interpret the origin and temporal evolution of these oscillations, it is useful to introduce a characteristic time scale for electron heating,
\begin{equation}
	\tau_{\mathrm{IB}}(t) \;\equiv\; \frac{U_e(t)}{Q_{\mathrm{abs}}(t)}
	\;=\; \frac{\frac{3}{2}\,n_e k_B T_e(t)}{\alpha_{\mathrm{IB}}(T_e)\,I(t)}
	\;\simeq\; \frac{T_e(t)}{dT_e/dt},
	\label{eq:tauIB}
\end{equation}
where $\tau_{\mathrm{IB}}$ is called the characteristic IB heating time,
$U_e(t) = \frac{3}{2}n_e k_B T_e(t)$ is the electron internal energy density,
{$Q_{\mathrm{abs}}(t)$ is the absorbed laser power density,
$\alpha_{\mathrm{IB}}(T_e)$ is the temperature dependent IB absorption coefficient per unit length (m$^{-1}$),}
$I(t)$ is the instantaneous laser intensity,
$n_e$ is the electron number density,
and $k_B$ is the Boltzmann constant.
Physically, $\tau_{\mathrm{IB}}$ represents the plasma response time to collisional laser heating,
that is, the time required for the electron temperature to undergo a significant fractional change due to IB absorption.
This timescale depends on both the laser and plasma parameters:
higher laser intensity or stronger electron–ion collisions (larger $\alpha_{\mathrm{IB}}$) shorten $\tau_{\mathrm{IB}}$, 
while higher electron temperatures, which reflect both reduced collisionality and increased internal energy, tend to produce a longer $\tau_{\mathrm{IB}}$.

The heating dynamics can be understood as a competition between the laser coherence time $\tau_c$ and the collisional heating time $\tau_{\mathrm{IB}}$, where the former characterizes the temporal fluctuation of the laser intensity and the latter governs the response of the electron temperature.
When $\tau_{\mathrm{IB}}$ is comparable to $\tau_c$, the plasma can respond rapidly enough to follow the short time intensity oscillations of the broadband field, producing visible fluctuations in $dT_e/dt$.
In the early stage of the present simulation, $\tau_{\mathrm{IB}}$ is estimated to be $\sim3$~ps, about ten times the laser coherence time ($\tau_{\mathrm{IB}}\!\sim\!10\tau_c$).
During this period, the heating rate difference $\Delta(dT_e/dt)$ exhibits pronounced oscillations that closely track the averaged intensity modulation on a comparable 2.8~ps timescale, as shown in figure~\ref{fig:Te_vs_t}(c).
This indicates that even when $\tau_{\mathrm{IB}}$ exceeds $\tau_c$ by an order of magnitude, the plasma still partially responds to the residual low frequency intensity variations that persist over several coherence intervals.
As the plasma heats up and $dT_e/dt$ decreases, $\tau_{\mathrm{IB}}$ increases rapidly and exceeds $200\tau_c$ after about 20~ps, over which the driving field intensity averaged across this timescale becomes nearly constant. 
At this stage, the plasma response is determined primarily by the time averaged intensity, and the oscillations in $\Delta(dT_e/dt)$ are strongly suppressed.
Hence, the electron plasma behaves as a low pass temporal filter with a cutoff determined by $\tau_{\mathrm{IB}}$: intensity components varying on timescales much shorter than $\tau_{\mathrm{IB}}$ are smoothed out by collisional energy exchange, leaving the net IB heating essentially insensitive to laser bandwidth.
The chosen diagnostic window of $2.8$~ps remains shorter than $\tau_{\mathrm{IB}}$ throughout the simulation, ensuring sufficient temporal resolution to capture the detailed evolution of $T_e(t)$.


Looking back at the overall trend, the broadband and monochromatic $T_e(t)$ curves gradually converge at later times, differing by less than $0.3\%$ at $t=35$~ps. 
This indicates that although temporal incoherence introduces short time fluctuations in $dT_e/dt$, the overall collisional absorption efficiency remains essentially unchanged. 
This behavior arises because both cases share the same central frequency and density ratio ($n_e/n_c$); hence, the absorption coefficient $\mu_c$ defined in equation~(\ref{eq:absorption}) is governed by the electron--ion collision frequency given in equation~(\ref{eq:ei_collision}). 
The latter depends primarily on the plasma state rather than on the spectral bandwidth. 
Therefore, the observed short term differences result from temporal intensity modulation rather than from intrinsic changes in the inverse bremsstrahlung cross section, explaining why the net heating efficiency and final equilibrium temperature are nearly identical for both monochromatic and broadband illumination.
{These findings further imply that, for radiation hydrodynamic simulations concerned with evolution on hundreds of picoseconds or longer~\cite{lei_2025, shestakov2000}, broadband illumination can be safely modeled using an equivalent monochromatic source, since bandwidth induced modulation does not leave lasting imprints on the collisional heating history.}


\section{Conclusions}

In this study, we have investigated the influence of broadband laser illumination on collisional inverse bremsstrahlung (IB) heating in Au plasmas.
Using one dimensional collisional particle-in-cell simulations, we isolated the effects of bandwidth while keeping the mean laser intensity, central frequency, and plasma parameters fixed.
The simulations were first benchmarked against classical absorption theory, confirming that the implemented collision model accurately reproduces the theoretical IB heating rate over a wide range of electron temperatures.

{
A detailed comparison of the electron temperature evolution was performed between the broadband and monochromatic cases.
In both scenarios, $T_e(t)$ rises rapidly at early times and then gradually approaches a saturated value.
The heating rate $dT_e/dt$ in the monochromatic case decreases almost monotonically, whereas the broadband case exhibits picosecond-scale oscillations that track the stochastic intensity fluctuations of the pump.
These oscillations, however, are transient and vanish when averaged over durations much longer than the coherence time.
The persistence or suppression of these fluctuations is governed by the relation between the characteristic IB heating time $\tau_{\mathrm{IB}}$ and the laser coherence time $\tau_c$:
when $\tau_{\mathrm{IB}}\!\gg\!\tau_c$, the plasma effectively averages over rapid intensity modulations, leading to a long term absorption efficiency that is indistinguishable from that of monochromatic illumination.
This indicates that, for radiation hydrodynamic simulations that typically evolve over hundreds of picoseconds or longer, the broadband laser can be effectively modeled as an equivalent monochromatic source of the same mean intensity.}


\section*{Acknowledgments}
This work was supported by the National Natural Science Foundation of China (Grant Nos. 12505268, 12275032, 12205021 and 12075227) and the China Postdoctoral Science Foundation (Grant No. 2024M764280).

\section*{Data availability statement}
The data cannot be made publicly available upon publication because the cost of preparing, depositing and hosting the data would be prohibitive within the terms of this research project. The data that support the findings of this study are available upon reasonable request from the authors.

\section*{References}

\bibliography{zotero-updating, wideband_clean}

\end{document}